\documentclass[]{spie}  

\usepackage{amsmath,amsfonts,amssymb}
\usepackage{graphicx}
\usepackage[colorlinks=true, allcolors=blue]{hyperref}
\usepackage[font=small]{caption}
\usepackage{subcaption}
\usepackage{booktabs}
\usepackage{tabu}
\usepackage{multirow}
\usepackage{float}
\usepackage{xcolor}

\title{Design and development of an ambient-temperature continuously-rotating achromatic half-wave plate for CMB polarization modulation on the POLARBEAR-2 experiment}

\author[a,b]{Charles A. Hill*}
\author[a]{Shawn Beckman*}
\author[a,c]{Yuji Chinone}
\author[a]{Neil Goeckner-Wald}
\author[c,d,e,f]{Masashi Hazumi}
\author[g]{Brian Keating}
\author[b]{Akito Kusaka}
\author[a,b,h]{Adrian T. Lee}
\author[g]{Frederick Matsuda}
\author[h]{Richard Plambeck}
\author[a,h]{Aritoki Suzuki}
\author[d,i]{Satoru Takakura}
\affil[a]{Department of Physics, University of California, Berkeley, CA 94720, USA}
\affil[b]{Physics Division, Lawrence Berkeley National Laboratory, Berkeley, CA 94720, USA}
\affil[c]{Kavli Institute for The Physics and Mathematics of The Universe (WPI), The University of Tokyo, Kashiwa, Chiba 277-8583, Japan}
\affil[d]{Institute of Particle and Nuclear Studies (IPNS), High Energy Accelerator Research Organization (KEK), Tsukuba, Ibaraki 305-0801, Japan}
\affil[e]{SOKENDAI (The Graduate University for Advanced Studies), Hayama, Kanagawa 240-0193, Japan}
\affil[f]{Institute of Space and Astronautical Science (ISAS), Japan Aerospace Exploration Agency (JAXA), Sagamihara, Kanagawa 252-5210, Japan}
\affil[g]{Department of Physics, University of California, San Diego, CA 92093-0424, USA}
\affil[h]{Radio Astronomy Laboratory, University of California, Berkeley, CA 94720, USA}
\affil[i]{Department of Earth and Space Science, Osaka University, Toyonaka, OSAKA 560-0043, Japan}

\authorinfo{*Charles Hill: chill@lbl.gov; +1 (763) 222-6359 \\ *Shawn Beckman: s.beckman@berkeley.edu; +1 (303) 819-2260}

\begin{document} 
\maketitle


\begin{abstract} 

We describe the development of an ambient-temperature continuously-rotating half-wave plate (HWP) for study of the Cosmic Microwave Background (CMB) polarization by the POLARBEAR-2 (PB2) experiment. Rapid polarization modulation suppresses 1/f noise due to unpolarized atmospheric turbulence and improves sensitivity to degree-angular-scale CMB fluctuations where the inflationary gravitational wave signal is thought to exist. A HWP modulator rotates the input polarization signal and therefore allows a single polarimeter to measure both linear polarization states, eliminating systematic errors associated with differencing of orthogonal detectors. PB2 projects a 365-mm-diameter focal plane of 7,588 dichroic, 95/150 GHz transition-edge-sensor bolometers onto a 4-degree field of view that scans the sky at $\sim$ 1 degree per second. We find that a 500-mm-diameter ambient-temperature sapphire achromatic HWP rotating at 2 Hz is a suitable polarization modulator for PB2. We present the design considerations for the PB2 HWP, the construction of the HWP optical stack and rotation mechanism, and the performance of the fully-assembled HWP instrument. We conclude with a discussion of HWP polarization modulation for future Simons Array receivers.

\end{abstract}



\section{INTRODUCTION} 
\label{sec:intro}

Precise characterization of the Cosmic Microwave Background (CMB) polarization anisotropies stands at the forefront of modern cosmology. Of particular importance is the parity-odd, divergence-free B-mode polarization pattern uniquely generated by primordial gravitational waves \cite{seljakGW, zaldarriagaAllSky, kamionkowskiCurl} and gravitational lensing \cite{huOkamoto, lewisChallinor}. The B-mode signal created by the primordial gravitational wave background (GWB) is thought to be the fingerprint of inflation \cite{guth, linde}---an epoch of exponential spatial expansion $\sim 10^{-30}$ seconds after the Big Bang---and peaks at degree angular scales. The B-mode signal created by gravitational lensing (GL) of parity-even E-modes into B-modes encodes information about large scale structure formation \cite{seljakGravPot, zaldProjMatDensity} and peaks at arcminute angular scales. The GWB remains undetected \cite{bicep2} while the GL signal is just beginning to be explored \cite{pb1BB, sptBB, actBB}. Therefore, there exists a wealth of B-mode physics yet to be harnessed, including the behavior of gravity at grand-unification energies \cite{kamionkowskiParticlePhys} and the impact of neutrinos on cosmological evolution \cite{abazajian}.




\subsection{The POLARBEAR-2 experiment} 
\label{sec:pb2}

The POLARBEAR-2 (PB2) receiver will observe the CMB polarization anisotropies from the Atacama Desert of Chile at 5,200 m altitude in 2017 \cite{nate, yuki}. The PB2 receiver mounts onto a telescope identical to the Huan Tran Telescope (HTT) \cite{tran} and observes at 95 GHz and 150 GHz simultaneously with a 4-degree field of view that scans the sky at $\sim$ 1 degree per second \cite{tomoOptics}. The 365-mm-diameter focal plane contains 1,897 dual-polarized, multi-chroic, planar-lithographed pixels that couple to the reimaging optics via an array of synthesized elliptical silicon lenslets \cite{pb2Det}. PB2's beam size of 3.5 arcmin (5.2 arcmin) at 150 GHz (95 GHz) and its 4.1 $\mu\mathrm{K}_{\mathrm{CMB}} \sqrt{\mathrm{s}}$ noise equivalent temperature make it well-suited to probe a wide range of angular scales. Therefore, PB2 aims to both characterize the GL signal and detect the GWB. 


\section{Half-wave plate polarization modulation} 
\label{sec:hwp}

Although PB2 achieves excellent white-noise performance, it is susceptible to low-frequency (or ``1/f'') noise induced by unpolarized atmospheric fluctuations leaking into polarization via instrumental effects \cite{errardATM}. Because atmospheric noise at the observation site is several orders of magnitude brighter than the B-mode signal and has  a 1/f knee $\sim$ 1 s \cite{absHWP}, even a small level of polarization leakage is detrimental. Additionally, PB2 is susceptible to instrumental effects that cause mismatch between orthogonal detectors, such as the ellipticity of the sinuous antenna beam \cite{pb2Det, tokiThesis}, and to differential response induced by the lenses and mirrors \cite{tranPol}. Because PB2 will achieve excellent statistical error, even small levels of these systematic errors can have a negative impact on experimental performance.

Continuous polarization modulation is a common technique to suppress 1/f noise \cite{maxipolHWP, absHWP, ebexHWP, actHWP, lbHWP, classVPM, piperVPM, bicepFRM}. A polarization modulator multiplies the input signal by a carrier wave set by the modulator's frequency, and the sky data is extracted via demodulation during analysis \cite{absHWP}. If the modulation frequency is well above the 1/f knee in temperature, white-noise sensitivity can be achieved on long timescales in polarization \cite{absHWP}. 

A rotating half-wave plate (HWP) mitigates systematic errors associated with mismatch between orthogonal detectors and the differential response of the optics to the orthogonal polarizations \cite{absHWP, absSyst, bryan}. A HWP spins the input polarization signal via rotation of a birefringent dielectric \cite{absHWP}, allowing a single polarimeter to measure both linear polarization states. This polarization rotation mitigates systematic effects due to beam asymmetry, differential gain, and dysfunctional detector pairs.

A rapidly-rotating HWP provides the benefit of both 1/f suppression and systematic error mitigation. When spinning rapidly, the HWP modulates the input polarization fast enough to reject atmospheric fluctuations while also allowing accurate separation of the sky signal from instrumental polarization \footnote{We note that the HWP only mitigates instrumental polarization due to elements between it and the detector.}. Given the multiple benefits offered by a rapidly-rotating HWP, we adopt it as the polarization modulator for PB2.


\subsection{POLARBEAR-1 HWP precedent} 
\label{sec:pb1hwp}

The predecessor to PB2 is the POLARBEAR-1 (PB1) experiment, which is currently in its fourth season of observation. The PB1 receiver is mounted on the HTT and contains an array of 637 dual-polarized pixels that observe at 150 GHz with 3.5-arcmin beams within a 2.2-degree field of view \cite{errardPB}. 

\begin{figure}
\centering
	\includegraphics[trim={1.5cm, 2.4cm, 1.5cm, 2.9cm}, clip, width=0.6\linewidth]{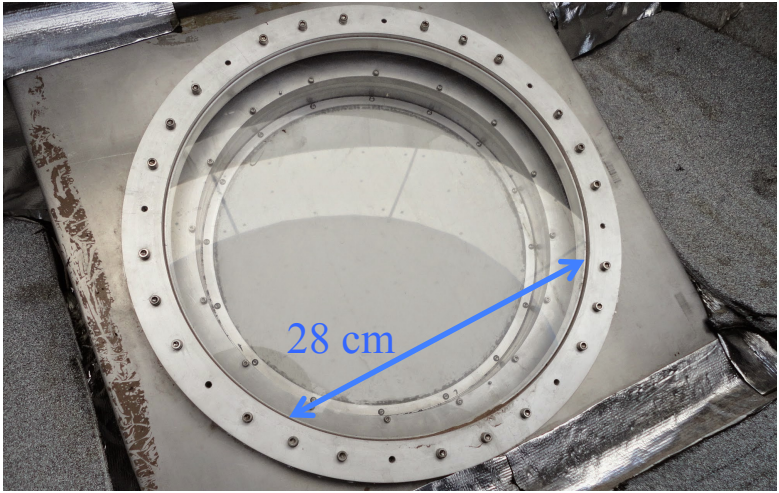}
	\caption{A photograph of the PB1 HWP deployed between the primary and secondary mirrors of the HTT. The HWP consists of a 280-mm-diameter sapphire window, AR coated with RT/duroid 6002 and covered by a Mylar sheet for weather protection. In this photo, the reflection of the primary mirror and guard ring can be seen on the Mylar sheet. \label{fig:pb1HWP}}
\end{figure}

In May 2014, PB1 deployed a continuously-rotating, ambient-temperature HWP, shown in Figure \ref{fig:pb1HWP}. The HWP consists of a single 280-mm-diameter, 3.1-mm-thick, $\alpha$-cut slab of sapphire from Crystal Systems \footnote{http://www.gtat.com/products-and-services-HEM-sapphire-for-optical.htm}, anti-reflection (AR) coated with 250-$\mu$m-thick RT/duroid 6002 from Rogers Corporation \footnote{https://www.rogerscorp.com/index.aspx}. The HWP is located near the primary focus between the primary and secondary mirrors of the HTT and rotates at $f_{\mathrm{m}} = 2$ Hz, hence modulating polarization at $4 f_{\mathrm{m}} = 8$ Hz \cite{absHWP}. 

PB1 HWP data analysis of a $\sim$ 1000 $\mathrm{deg}^{2}$ sky patch shows that the HWP provides significant suppression of atmospheric 1/f power in the polarization channels \cite{satoru}. Because PB1 operates at the same site and with an identical telescope design as PB2, we find the positive experiences of the PB1 HWP to be both encouraging and informative as we move towards deployment of a HWP for PB2.

\subsection{The achromatic HWP} 
\label{sec:ahwp}

Because PB2 observes in two frequency bands using a single receiver, we adopt a Pancharatnam \cite{ahwp} achromatic HWP (AHWP) to modulate 95 GHz and 150 GHz simultaneously. The PB2 AHWP consists of three identical 3.75-mm-thick sapphire plates with central frequency $\approx$ 120 GHz stacked such that the relative orientations between the plates' crystal axes optimizes modulation bandwidth \cite{savini}. 

\begin{figure}[H]
\begin{subfigure}{\textwidth}
  	\centering
  	\includegraphics[trim={1.5cm, 3.5cm, 1.5cm, 5.0cm}, clip, width=0.7\textwidth]{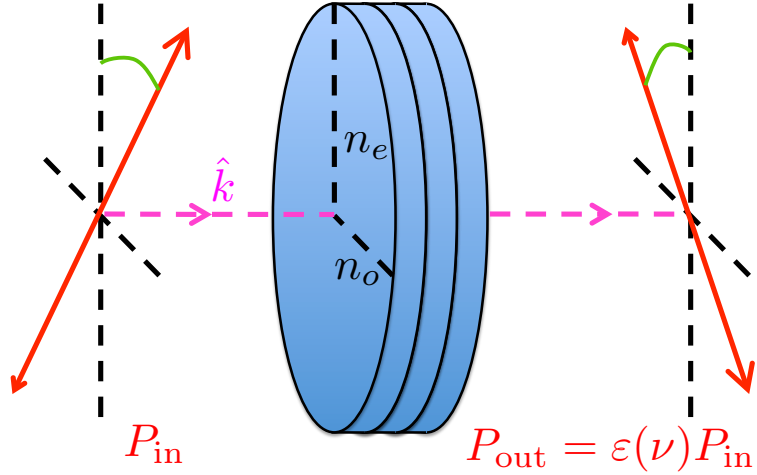}
  	\caption{\label{fig:HWPOper}}
	\par\vfill
\end{subfigure}
\begin{subfigure}{0.5\textwidth}
	\centering
  	\includegraphics[trim={3.5cm, 1.0cm, 3.5cm, 2.5cm}, clip, width=\textwidth]{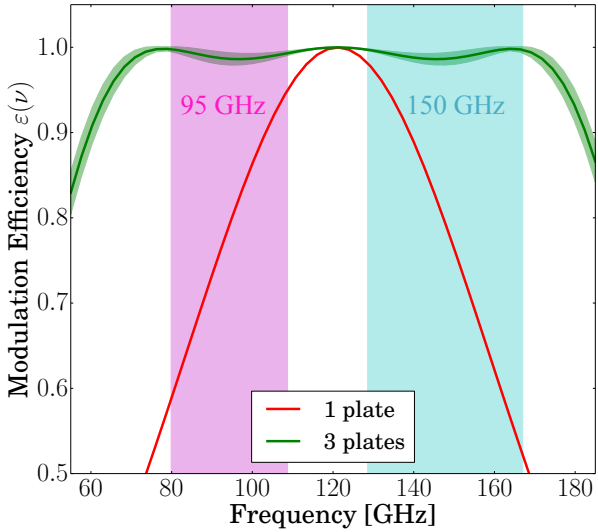}
  	\caption{\label{fig:modEff}}
\end{subfigure}
\begin{subfigure}{0.5\textwidth}
	\centering
  	\includegraphics[trim={3.5cm, 1.0cm, 3.5cm, 2.5cm}, clip, width=\textwidth]{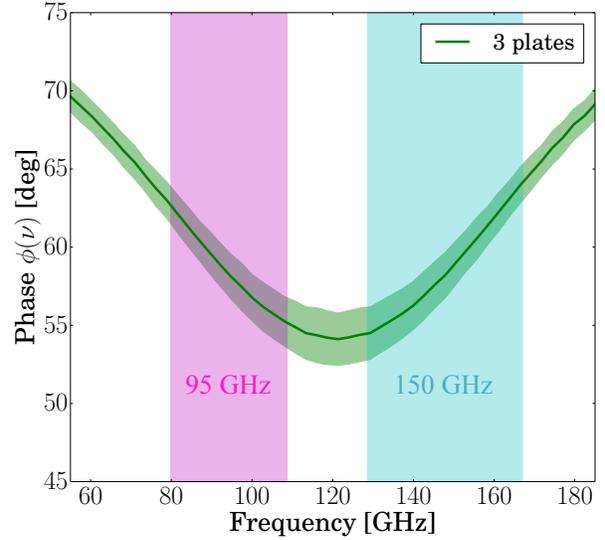}
  	\caption{\label{fig:phase}}
\end{subfigure}
\caption{An overview of the AHWP operating principle. \ref{fig:HWPOper} shows that an AHWP rotates input linearly-polarized light by twice the polarization angle with respect to the sapphire crystal axes plus a frequency-dependent phase and with a frequency-dependent modulation efficiency. The ordinary and extraordinary crystal axes of the skyward-most sapphire plate are labelled $n_{\mathrm{o}}$ and $n_{\mathrm{e}}$, respectively. \ref{fig:modEff} and \ref{fig:phase} show the modulation efficiency and phase versus frequency for a single and three-stack design. Note that the phase of the single HWP is zero for all frequencies. The green shaded regions are associated with tolerances on the relative alignment of the stack as defined in Table \ref{table:plateOrientations}. The assumed thickness and indexes of the sapphire are given in Table \ref{table:measOptics}. The magenta and cyan regions show the PB2 frequency bands. \label{fig:HWPTheory}}
\end{figure}

The AHWP operating principle is shown in Figure \ref{fig:HWPOper}. Incoming linearly-polarized light is rotated by twice its polarization angle with respect to the sapphire crystal axes $\theta_{\mathrm{in}}$, plus a frequency-dependent phase $\phi(\nu)$ \cite{ahwp}
\begin{equation}
	\Delta \theta = 2 \big[ \theta_{\mathrm{in}} - \phi(\nu) \big] \, .
\label{eq:rot}
\end{equation}
Given linear-polarization Stokes parameters Q and U, we define the total linear polarization as
\begin{equation}
	P = \sqrt{Q^{2} + U^{2}}
\label{eq:polFrac}
\end{equation}
and the HWP modulation efficiency as
\begin{equation}
	\varepsilon = \frac{P_{\mathrm{out}}}{P_{\mathrm{in}}} \, ,
\label{eq:modEff}
\end{equation}
where $P_{\mathrm{out}}$ ($P_{\mathrm{in}}$) is the total linear polarization of the output (input) light. 

AHWP modulation efficiency and phase versus frequency \cite{tomo} are plotted in Figures \ref{fig:modEff} and \ref{fig:phase}. The assumed AHWP plate angles as well as the PB2-band-averaged modulation efficiency and phase are listed in Table \ref{table:plateOrientations}. PB2 modulation efficiency improves from 78\% to 99\% via the implementation of a Pancharatnam design over a single-plate design. We note that the overall phase offset introduced by the AHWP design will be calibrated out as a global rotation and therefore is not meaningful \cite{keating}. Instead, it is the phase difference between the 95 GHz and 150 GHz bands that requires careful measurement and calibration \cite{tomo, bao}.

\begin{table}[H]
\centering
	\begin{tabu}{| c | c | c | c | c | c |}
	\hline
	\multirow{2}{*}{Configuration} & Plate Orientations & $\varepsilon(95 \; \mathrm{GHz})$ & $\phi(95 \; \mathrm{GHz})$ & $\varepsilon(150 \; \mathrm{GHz})$ & $\phi(150 \; \mathrm{GHz})$ \\
	 & [deg] & [\%] & [deg] & [\%] & [deg] \\
	\hline
	\hline
	Single HWP & [$0.0$] & $78.4$ & $0.0$ & $78.0$ & $0.0$ \\
	\hline 
	AHWP & [$0.0$, $54.0 \pm 1.0$, $0.0 \pm 1.4$] & $98.9 \pm 0.6$ & $58.4 \pm 1.3$ & $99.1 \pm 0.5$ & $58.6 \pm 1.3$ \\	
	\hline
	\end{tabu}
\caption{Single and three-stack HWP relative crystal-axes orientations with their associated modulation efficiencies and phases integrated across the PB2 frequency bands. We assume a one-degree alignment tolerance between adjacent plates that propagates as uncorrelated angle uncertainty during stack construction. Note that the global AHWP phase offset is not meaningful; instead, the important quantity is the phase difference between the 95 GHz and 150 GHz bands. \label{table:plateOrientations}}
\end{table} 



\section{PB2 HWP Overview} 
\label{sec:req}

PB2 presents unique challenges to the implementation of a polarization modulator: a large optical throughput, a broad bandwidth, and a photon-noise-limited detector array whose sensitivity relies on low levels of optical loading. 

\begin{figure}
\centering
	\includegraphics[trim={1.5cm, 4.5cm, 1.5cm, 2.5cm}, clip, width=0.8\linewidth]{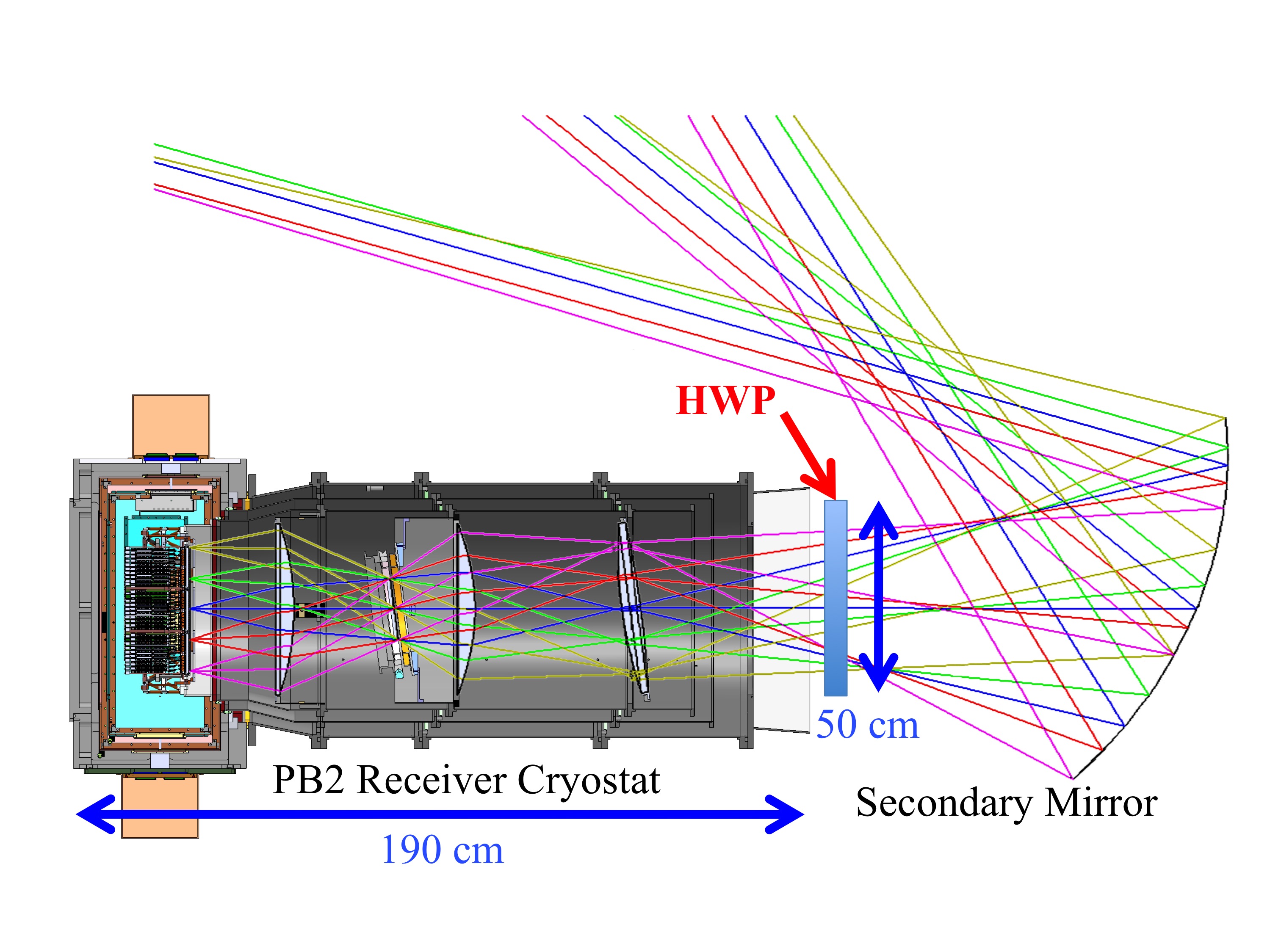}
\caption{The primary drivers for the HWP design are PB2's large optical throughput, broad bandwidth, and stringent optical loading requirements. The HWP is located at ambient temperature near the receiver cryostat window and must be $\approx$ 470 mm in diameter to cover all rays, including margin for diffraction effects.\label{fig:HWPPos}}
\end{figure}

Figure \ref{fig:HWPPos} shows a Zemax ray trace of the PB2 receiver and the location of the HWP in the optical path. The HWP will be located just beyond the cryostat window and must be $\geq$ 470 mm in diameter to cover the beam from every detector on the focal plane, including margin for diffraction effects. This requires a birefringent substrate of diameter $\gtrsim$ 500 mm to accommodate clamping and alignment tolerances.

Because PB2 detectors are designed to be photon-noise limited \cite{nate}, PB2 sensitivity is adversely affected by HWP thermal emission and scattering to ambient temperature. The achromatic design is three times the thickness of a single HWP and has multiple layers of AR coating, necessitating that the PB2 HWP be composed of very-low-loss materials. Additionally, due to its location outside the cryostat, the HWP does not see a cryogenic environment across its $2\pi$ sr scattering volume, necessitating that the HWP consist of AR layers that achieve small reflection across both the 95 GHz and 150 GHz bands.


The PB2 HWP not only poses optical challenges but mechanical ones as well. In order to reject 1/f noise, we must rotate the HWP quickly enough to achieve white-noise sensitivity in our polarization measurements while also minimizing vibrational coupling to the telescope and maintaining adequate rotational stability. We use positive experiences from PB1 HWP analysis to set the PB2 HWP rotation frequency at $f_{\mathrm{m}} =$ 2 Hz hence modulating polarization at a carrier frequency of 4$f_{\mathrm{m}} =$ 8 Hz. This speed corresponds to $\sim 10^{8}$ HWP revolutions during three years of PB2 operation and poses challenges to the design of the drive mechanism.



\section{Optical stack} 
\label{sec:optStack}

The PB2 HWP consists of three 512-mm-diameter, 3.75-mm-thick, $\alpha$-cut sapphire plates fabricated via the advanced heat-exchange method at Guizhou Haotian Optoelectronics Technology \footnote{http://www.ghtot.com/} (GHTOT). The plates are cut and ground with an $\alpha$-plane alignment of $\pm$ 2 degrees, a surface parallelism of $\pm$ 100 $\mu$m, and a surface roughness of 0.5 $\mu$m RMS. 

The AR coating consists of 380-$\mu$m-thick sheets of high-density polyethylene (HDPE)  from New Process Fibre \footnote{http://www.newprocess.com/} and 270-$\mu$m-thick sheets of RO3006 circuit board laminate from Rogers Corporation. The AR layers are pressed onto the sapphire via atmospheric pressure, a technique meant to eliminate thermal emission associated with glue layers and to avoid potential AR delamination during observations. The vacuum module is detailed in Section \ref{sec:design}. 

Table \ref{table:measOptics} presents the measured parameters of the optical stack. The AR indexes were obtained using a Fourier Transform Spectrometer (FTS) (see Figure \ref{fig:FTS}), the sapphire indexes using an FTS with aligned wiregrids on either side of the sample, and the loss tangents using the thermal emission apparatus described in Section \ref{sec:sappLoss}.

\begin{table}
	\centering
	\begin{tabu}{| c | c | c | c |}
	\hline
	Stack Element & Thickness [mm] & Index of Refraction & Loss Tangent [$\mathrm{10^{-4}}$] \\
	\hline
	\hline
	Top AR Layer: HDPE & $0.38 \pm 0.04$ & $1.55 \pm 0.01$ & $0.5 \pm 1.0$ \\
	\hline
	Bottom AR Layer: RO3006 & $0.27 \pm 0.02$ & $2.52 \pm 0.01$ & $56.5 \pm 2.7$ \\
	\hline
	Sapphire Ordinary Axis & \multirow{2}{*}{$3.75 \pm 0.01$} & $3.05 \pm 0.03$ & $0.1 \pm 1.3$ at 95 GHz \\
	\cline{1-1} \cline{3-3}
	Sapphire Extraordinary Axis & & $3.38 \pm 0.03$ & $1.1 \pm 1.3$ at 150 GHz \\
	\hline
	\end{tabu}
\caption{The measured ambient-temperature thickness, index of refraction, and loss tangent for each layer in the HWP optical stack. Indexes are measured using a Fourier Transform Spectrometer (Figure \ref{fig:FTS}) and loss tangents are measured using a thermal emission apparatus (Figure \ref{fig:tea}). The sapphire loss tangent is measured at 100, 110, 220, 230, and 240 GHz, is averaged over the ordinary and extraordinary axes, and is interpolated to the PB2 frequencies, as described in Section \ref{sec:sappLoss}. \label{table:measOptics}}
\end{table}


\subsection{Sapphire evaluation} 
\label{sec:sappLoss}

PB2 HWP emission is quite sensitive to dielectric loss in the birefringent substrate, as it consists of $3.75$ mm $\times$ 3 $=$ 11.25 mm of sapphire with an axis-averaged refractive index of $\approx$ 3.2. Because PB2 sensitivity is a strong function of HWP emission, a precise measurement of sapphire loss tangent (tan $\delta$) in the PB2 frequency range is critical to the HWP design process. 

\begin{figure}[h!]
\centering
\begin{subfigure}{.5\textwidth}
  	\centering
  	\includegraphics[trim={1.5cm, 2.5cm, 1.5cm, 2.5cm}, clip, width=\linewidth]{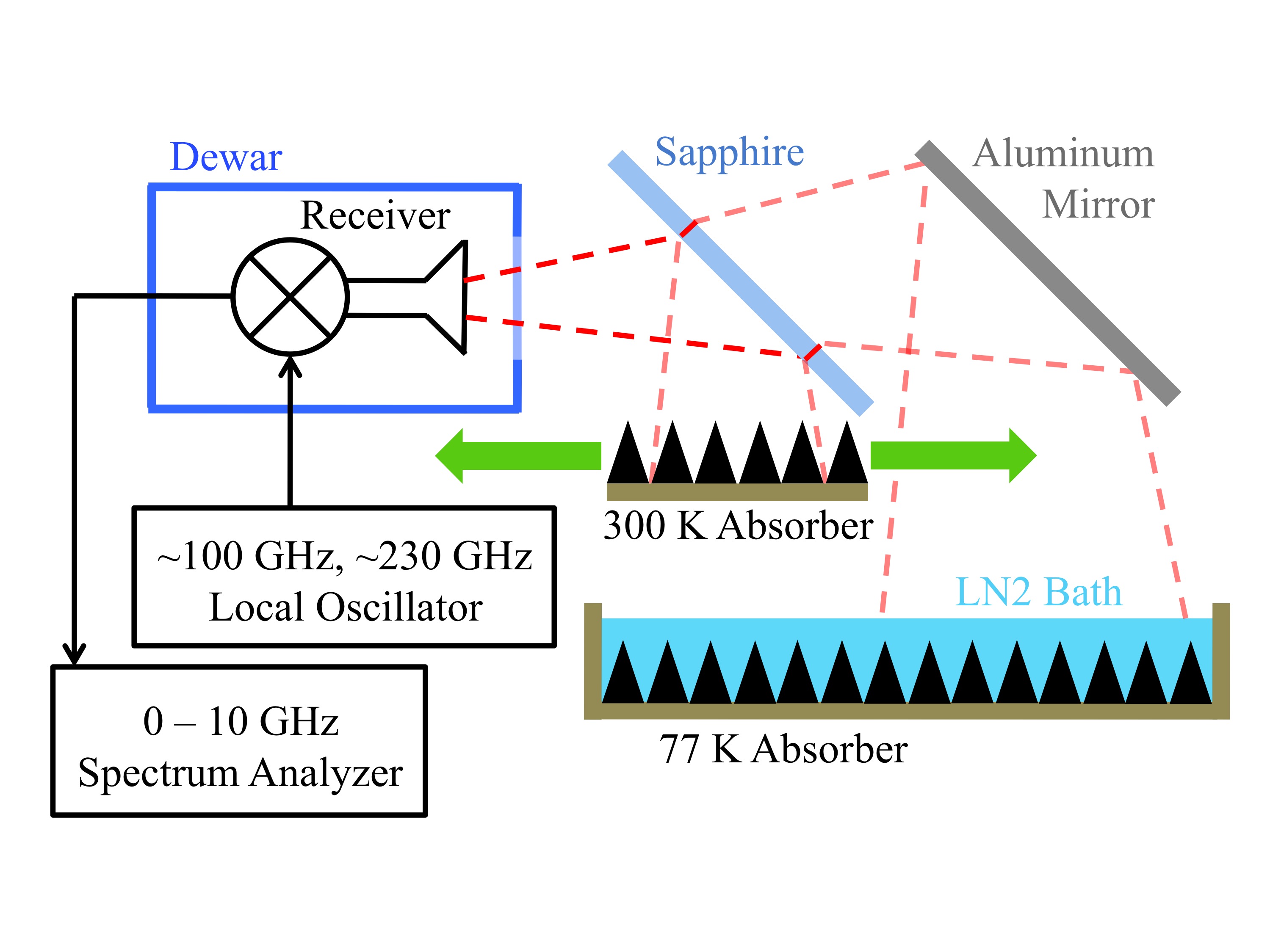}
   	\caption{\label{fig:tea}}
\end{subfigure}%
\begin{subfigure}{.5\textwidth}
  	\centering
  	\includegraphics[trim={1.5cm, 1.0cm, 1.5cm, 2.5cm}, clip, width=\linewidth]{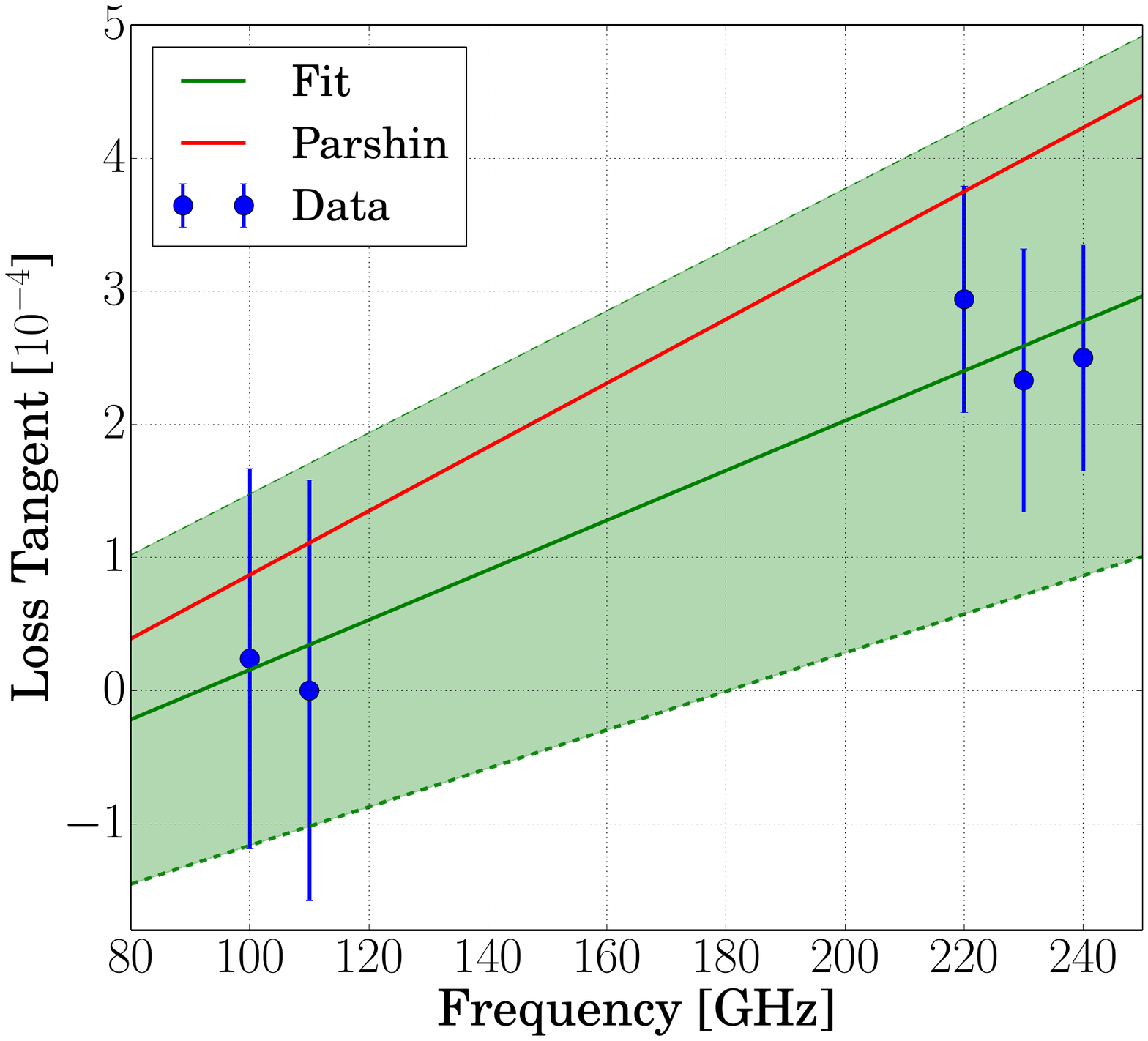}
  	\caption{\label{fig:sappLT}}
\end{subfigure}
\caption{PB2 HWP sapphire tan $\delta$ is measured using the TEA shown schematically in \ref{fig:tea}. Thermal emission from the sapphire slab is measured over a 77 K background using a heterodyne receiver. A series of configurations involving 300 K absorber are used to calibrate receiver gain, receiver noise, reflection from the sapphire, and transmission through the sapphire. \ref{fig:sappLT} shows a measurement of the IF-band-averaged, axes-averaged GHTOT sapphire tan $\delta$ at 5 LO frequencies with error bars that account for both statistical and systematic uncertainty in blue, a linear fit to those data points in green, and data presented in Parshin et al. \cite{parshin} in red. \label{fig:Sapphire}}
\end{figure}

Sapphire is expected to have tan $\delta \sim 10^{-4}$ at 150 GHz \cite{lamb} and 300 K, making an emissivity measurement of 3.75-mm-thick slabs challenging \footnote{CMB optical engineers traditionally test the loss of small, thick samples before purchasing the full-scale optic. However, sapphire quality is sensitive to the properties of its specific growth and the location within the boule. Additionally, it is expensive to purchase small, thick pieces from growths intended for large-diameter slabs. Therefore, we choose to measure and ``certify'' the 512-mm-diameter, 3.75-mm-thick pieces that comprise the deployment-ready HWP instrument.}. To meet this challenge, we construct the thermal emission apparatus (TEA) shown schematically in Figure \ref{fig:tea}. 

The TEA measures the temperature increase over a 77 K background due to thermal emission from a dielectric slab. We use a 100 GHz/300 GHz heterodyne receiver previously used on the Combined Array for Research in Millimeter-wave Astronomy (CARMA). The receiver features horn-coupled SIS (superconducting-insulating-superconducting) mixers cooled to 4 K, fed by a tunable local oscillator (LO) to produce a 1-10 GHz intermediate-frequency (IF) band that is digitized with a spectrum analyzer. 

Because the sapphire is not AR coated during this evaluation, it is important that radiation reflected from the slab's front surface terminates on 77 K. Thus, as shown in Figure \ref{fig:tea}, the sapphire is mounted at a 45-degree angle to the receiver in front of an aluminum sheet ($\lesssim$ 1 $\mu$m RMS surface roughness) also mounted at 45 degrees. Both the sapphire and the mirror reflect to pyramidal absorber submerged in a liquid nitrogen bath. The sapphire is on a rotating stage, and the receiver is optimized for p-polarization, so as to maximize signal transmission through the slab.

Measurements were made at LO frequencies of 100, 110, 220, 230, and 240 GHz. At each frequency we made a series of five measurements listed in Table \ref{table:teaConfigs} in order to solve for the receiver noise temperature and gain as well as the sapphire reflection coefficient, transmission coefficient, and emissivity.

\begin{table}
	\centering
	\begin{tabu}{| c | c | c | c |}
	\hline
	Sapphire & Reflected & Transmitted & Purpose \\
	\hline
	\hline
	- & - & 77 K & Calibrate receiver noise temp \\
	\hline
	- & - & 300 K & Calibrate receiver gain \\	
	\hline
	Y & 300 K & 77 K & Calibrate reflection \\	
	\hline
	Y & 77 K & 300 K & Calibrate transmission \\ 
	\hline
	Y & 77 K & 77 K & Measure emission \\
	\hline
	\end{tabu}
\caption{TEA configurations used to characterize the GHTOT sapphire tan $\delta$. The first column shows whether the sapphire is present, the second which temperature the sapphire reflects to, the third which temperature the mirror reflects to, and the fourth the configuration's purpose. \label{table:teaConfigs}}
\end{table}

Several factors complicate the calculation of tan $\delta$. First, the sapphire slab acts as a Fabry-Perot etalon, creating a series of transmission peaks with $\sim$ 13 GHz spacing; absorption, and hence thermal emission, in the slab is maximized at these frequencies and minimized at the reflection peaks between them. Second, measuring this pattern is complicated by the double sideband receiver, which folds signals above and below the LO frequency into a single IF band. Finally, the principle axes of the sapphire were not accurately known at the time of these measurements. Simulations based on analytic techniques \cite{tomCode, hou} show that averaging the data across the IF band and over a range of sapphire azimuth angles incurs systematic error that is smaller than the noise in our measurement. Therefore, we present the band-averaged, sapphire-angle-averaged tan $\delta$ values in Figure \ref{fig:sappLT}. The measured GHTOT sapphire tan $\delta$ is consistent with the literature value \cite{parshin} to within our 1$\sigma$ uncertainty of $\sim10^{-4}$.

To estimate the sapphire tan $\delta$ in the PB2 frequency bands, we linearly interpolate the measured tan $\delta$. The results of the tan $\delta$ interpolation to 95/150 GHz are shown in Table \ref{table:measOptics}. We find that the absorption in the GHTOT sapphire stack is smaller than the absorption in the RO3006 AR layers in the PB2 frequency range; therefore, we adopt GHTOT sapphire as the birefringent dielectric for the PB2 HWP.


\section{Mechanical assembly} 
\label{sec:design}

\begin{figure}
\centering
	\includegraphics[trim={1.5cm, 1.5cm, 1.5cm, 1.5cm}, clip, width=0.7\textwidth]{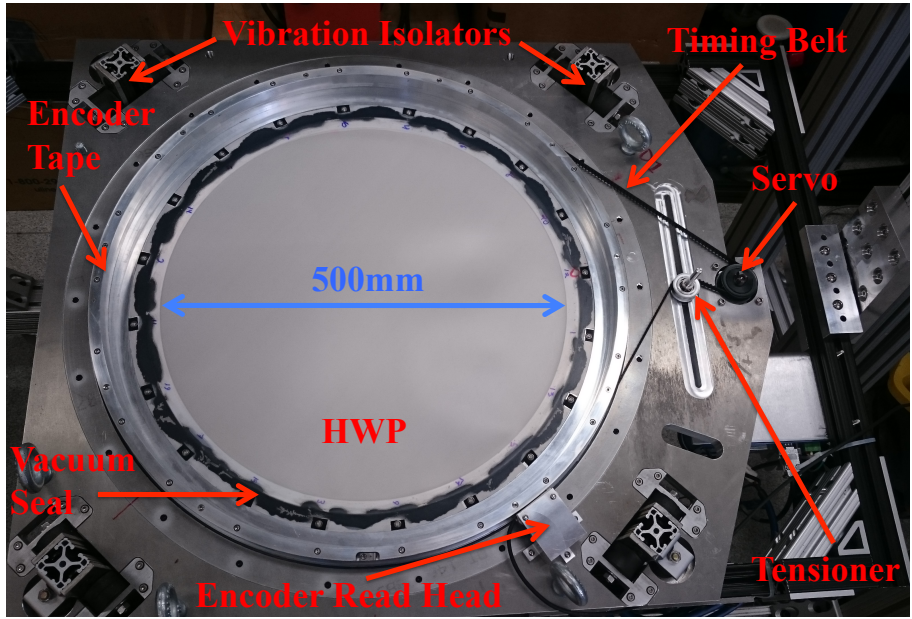}
	\caption{The fully-assembled HWP on its teststand at Berkeley. Important aspects of the design include a vacuum module to apply the AR coating, a servo motor coupled to a timing belt for modularity and rotational stability, and vibration damping mounts to mechanically isolate the HWP from the rest of the PB2 instrument. \label{fig:assem}}
\end{figure}

The HWP mount and rotation stage must spin $\sim$ 500-mm-diameter optics at 2 Hz for more than 100 million revolutions during three years of PB2 observation. Furthermore, the HWP assembly must minimize vibrational coupling to the telescope and receiver, maintain adequate rotational stability, be flexible during in-field optical alignment, and be robust against mechanical complications. In this section, we describe the key features of the PB2 HWP mechanical assembly \footnote{Large-diameter parts for the HWP assembly were fabricated by Jeff Tiedeken at the Monkey Likes Shiny machine shop in Berkeley.} shown in Figure \ref{fig:assem}. 

\subsection{Mount} 
\label{sec:mount}

The HWP mount is designed to maximize in-field flexibility. The mount attaches to the telescope boom via a series of $\mathrm{80/20^{\textregistered}}$ rails and rotational stages, allowing for fine control along three translational and two rotational axes. Additionally, the mount is capable of positioning the HWP either in front of the cryostat window or between the primary and secondary mirrors (just as in PB1). This locational flexibility allows for in situ evaluation of HWP-related aberrations, instrumental polarization, and beam coverage.

In order to ensure that HWP vibrations are isolated from the telescope and vice versa, the assembly is coupled to the telescope using a series of twelve independent rubber sandwich mounts oriented both tangentially and axially to the HWP rotation axis. To isolate the sapphire from vibrations in the bearing, the optical stack is clamped by a thin rubber gasket.


\subsection{Rotation mechanism} 
\label{sec:rot}

For the rotation stage, we utilize a 635-mm-diameter matched-pair thin-section ball bearing from SilverThin Bearings \footnote{http://www.silverthin.com/}. This bearing is a scaled-up version of that used on PB1 and is chosen for its proven success at the observation site. The bearing is preloaded to meet product specifications and is lubricated with a low-temperature-compatible grease to accommodate on-site weather conditions. The bearing races are stainless steel, and the bearing clamps are 7075 aluminum to minimize weight. We are in the midst of performing bearing stress and thermal tests to ensure robustness against premature wear and seizing.

The drivetrain is a 400 W, 60 mm AC servo from Applied Motion Products \footnote{http://www.applied-motion.com/}. This motor accommodates our estimated peak and continuous torque requirements with a safety factor of $\approx$ 3 and avoids electrical switching noise present in comparable stepper motors. The servo is tuned to provide loose feedback, keeping the motor from overheating and utilizing the rotor's large moment of inertia to maintain stable rotation. The drivetrain is connected to the rotor via a Kevlar-reinforced timing belt and a commercial tensioner pulley, a system that allows for easy installation of replacement parts at the observation site.

The HWP will be enclosed in the weatherproofed telescope boom, making it safe from winds and snow but not from fluctuations in ambient temperature. As a result, all HWP components are rated to -20 degrees Celsius.


\subsection{AR vacuum module} 
\label{sec:vacuum}

\begin{figure}
\centering
	\includegraphics[trim={0.0cm, 17.5cm, 1.5cm, 19.0cm}, clip, width=0.8\textwidth]{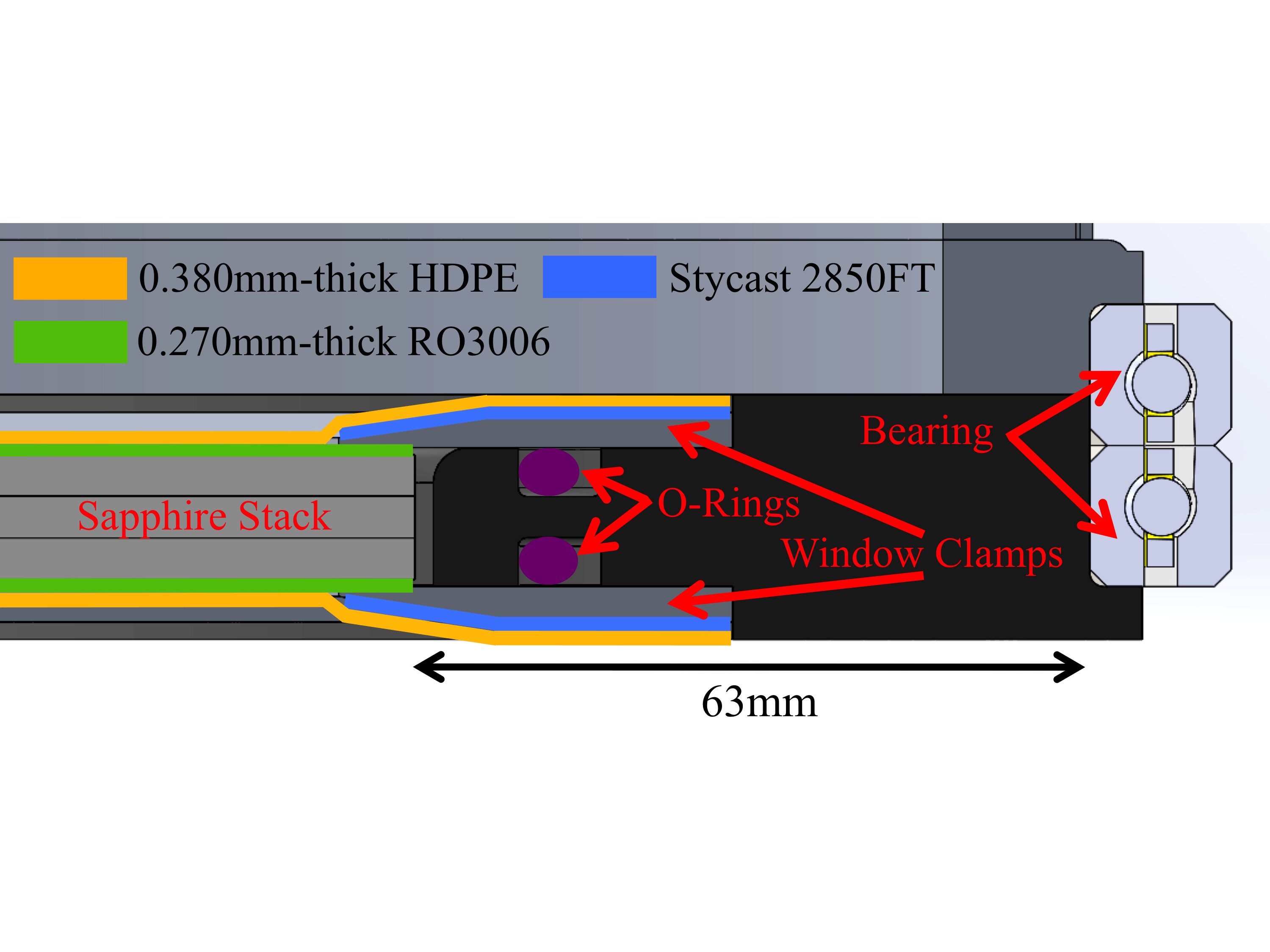}
	\caption{A cross-sectional view of the vacuum module. The HDPE AR layers are glued to aluminum annuli which mate to o-ring seals on the rotor. When the module is evacuated, atmosphere presses the AR layers onto the sapphire, hence eliminating the need for glue layers. \label{fig:crossSec}}
\end{figure}

In an effort to minimize HWP thermal emission and avoid AR delamination, we eliminate glue layers by ``adhering'' the optical stack via pressure gradients. A SolidWorks cross-sectional view of the AR vacuum module is shown in Figure \ref{fig:crossSec}. The HDPE AR layer is glued via Stycast 2850FT to an aluminum annulus, creating a circular ``window clamp''. The aluminum annuli are screwed onto Buna-N o-rings such that the RO3006 sheets and the sapphire plates sit between the HDPE windows. We then evacuate the inner chamber via a threaded port coupled to a T-valve, creating a vacuum space that allows atmospheric pressure to secure the optical stack. After disconnecting the pump, the module is mounted on the rotor such that the sapphire is free to rotate while under vacuum.


\subsection{Encoder and readout} 
\label{sec:readout}

\setcounter{footnote}{0}

\begin{figure}
\centering
	\includegraphics[trim={1.0cm, 3.5cm, 1.5cm, 5.0cm}, clip, width=0.7\textwidth]{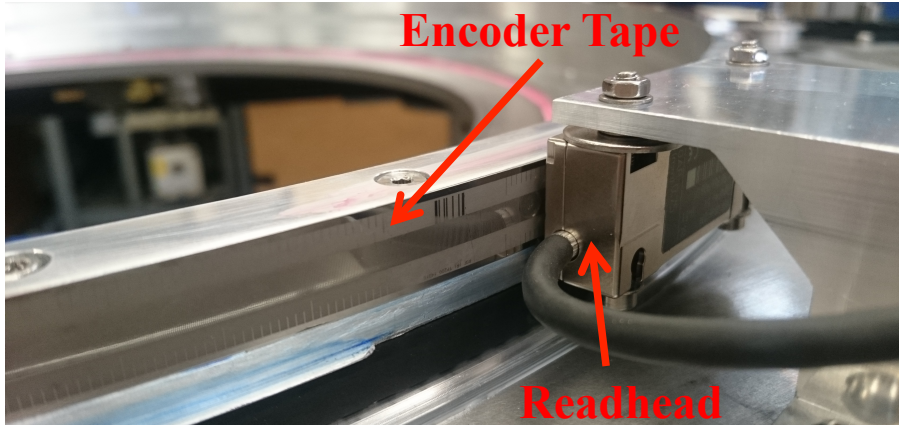}
	\caption{A photograph of the encoder tape and readhead. Readhead gap tolerance is $+0.4$ mm/$-0.2$ mm, necessitating $<$ 0.03\% ellipticity for the 635-mm-diameter encoder tape mounting ring. \label{fig:encoder}}
\end{figure}

To suppress jitter in the HWP angle reconstruction that can add noise during the analysis process, we must achieve a timing resolution much finer than 10 ms, which would be the digitization noise level if we naively read out the encoder at the 100 Hz detector sampling rate. We utilize a commercial optical encoder from RSF Elektronik \footnote{http://www.rsf.at/en/} that consists of a steel tape with 10,000 reflective lines and an infrared readhead, as shown in Figure \ref{fig:encoder}. The encoder tape is mounted on a 636-mm-diameter aluminum ring such that its surface maintains a radial distance of 0.75 mm $+0.4$ mm/$-0.2$ mm from the readhead during HWP rotation. This system provides 6.5 arcsec resolution via 4-times interpolation, which is fine enough to inject negligible noise into our demodulated data.

To read out the encoder, we use an Arduino Leonardo ETH microcontroller. The controller samples the encoder signal at 40 kHz and averages the data down at the 100 Hz detector sampling rate to achieve high timing resolution. The Arduino then compiles a UDP packet containing the HWP angle and a universal timestamp (IRIG), and this packet is sent to a central computer where it is verified and paired with a matching data packet. Using a microcontroller eliminates the need for a HWP-dedicated computer and allows for cheap, quick replacement in the case of failure at the site.


\section{HWP validation} 
\label{sec:hwpValid}

We report on the AR, polarization, and mechanical validations of the fully-constructed HWP shown in Figure \ref{fig:assem}. These tests are primers for PB2 full-system integration later this year.

Validation of the HWP optical stack was performed using a test pixel consisting of a transition-edge-sensor bolometer fed by a lenslet-coupled sinuous antenna \cite{pb2Det}. The detector is cooled to 250 mK in an optical Infrared Labs dewar \cite{pb2Det} and is read out using a DC superconducting quantum interference device (SQUID) amplifier from Quantum Design \footnote{http://www.qdusa.com/products/laboratory-squids.html}.


\subsection{AR coating validation} 
\label{sec:arValid}

\begin{figure}
\centering
\begin{subfigure}{.49\textwidth}
  	\centering
  	\includegraphics[trim={1.5cm, -2.0cm, 1.5cm, 1.5cm}, clip, width=\linewidth]{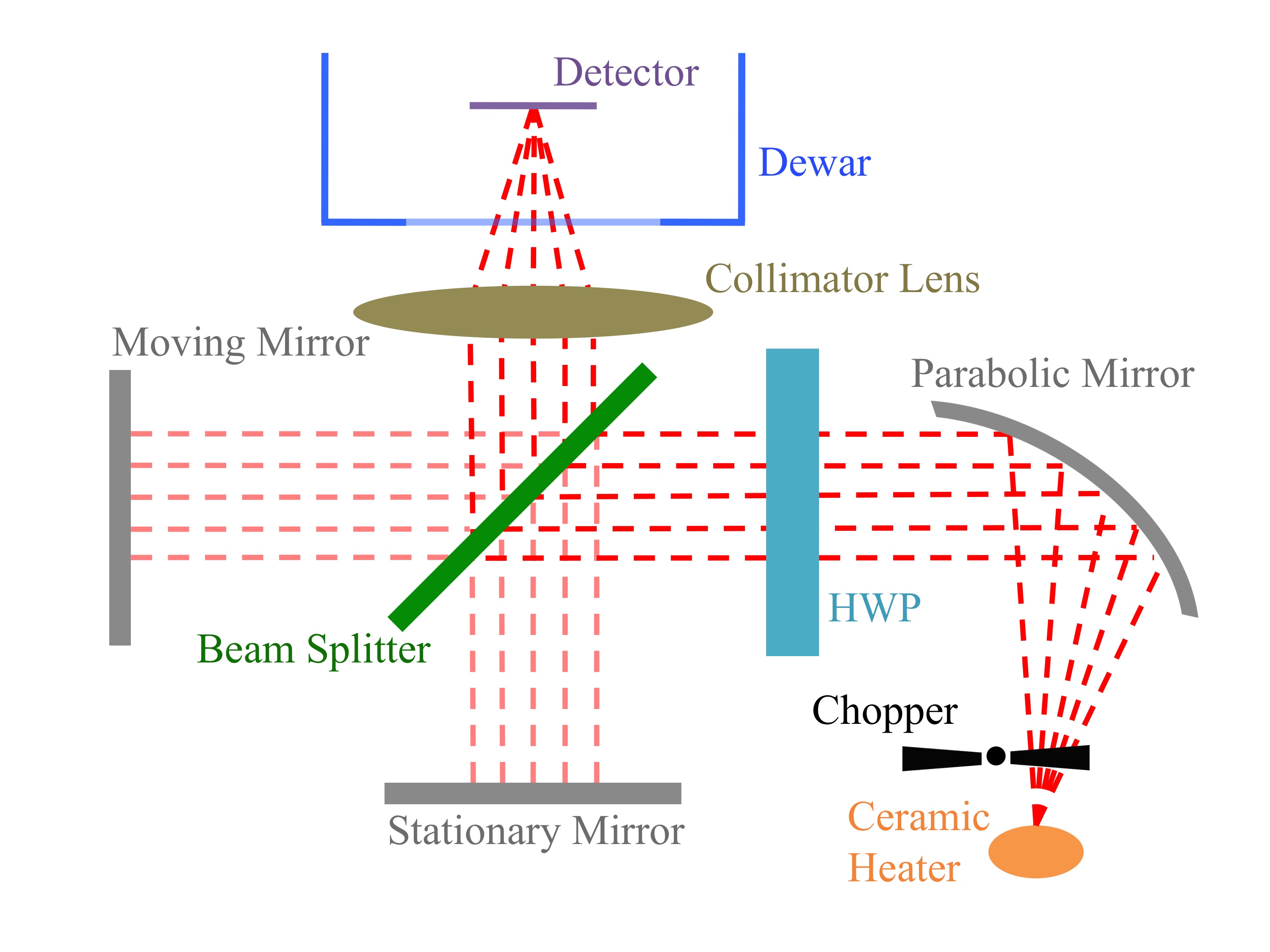}
   	\caption{\label{fig:FTS}}
\end{subfigure}%
\begin{subfigure}{.51\textwidth}
  	\centering
  	\includegraphics[trim={1.5cm, 1.3cm, 1.5cm, 1.5cm}, clip, width=\linewidth]{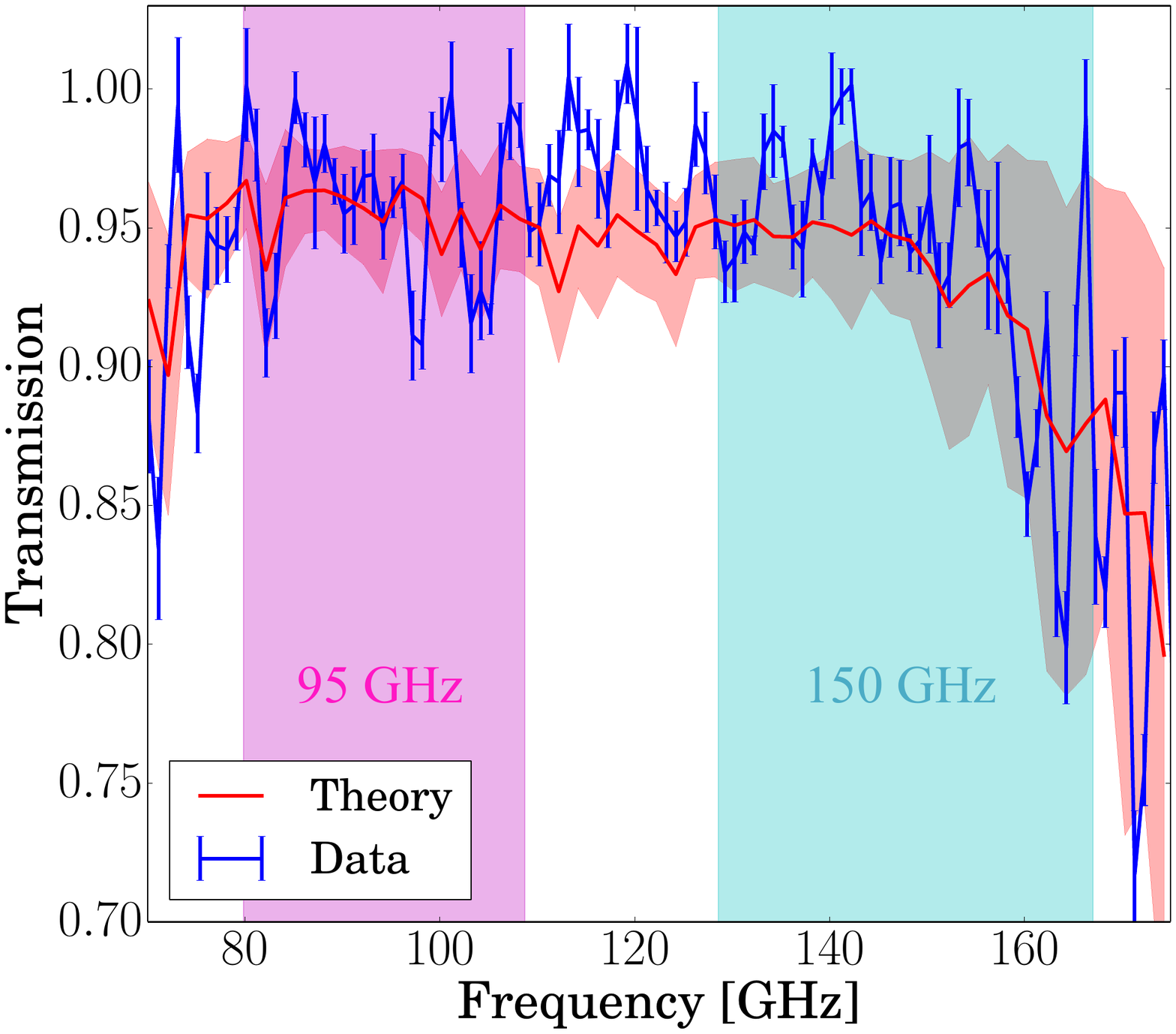}
  	\caption{\label{fig:hwpBandpass}}
\end{subfigure}
\caption{Measurement of the transmission through the PB2 HWP. \ref{fig:FTS} shows a cartoon of the FTS apparatus used to measure the HWP transmission. Signal from a temperature-modulated source is collimated, travels through the HWP and into the FTS, and is focused onto a cryogenic test pixel. \ref{fig:hwpBandpass} shows the result of the measurement in blue and the theoretical transmission curve constructed from the parameters listed in Table \ref{table:measOptics} in red. The measurement and the theory agree to within 1$\sigma$ uncertainty. \label{fig:bandpass}}
\end{figure}

To validate the HWP AR performance, we utilize an FTS coupled to a broadband detector, as shown schematically in Figure \ref{fig:FTS}. Signal from a temperature-modulated source is collimated by an off-axis parabolic mirror. This beam travels through the HWP, is split between a pistoned and fixed path by a 250-$\mu$m-thick Mylar film, is recombined by the same splitter, and focuses onto the detector via an ultra-high-molecular-weight polyethylene (UHMWPE) collimator lens. The SQUID output is sent to a lock-in amplifier whose value is integrated for 0.5 s at each moving-mirror position. The mirror step size and dynamic range give a 1 GHz resolution over a 300 GHz bandwidth. 

We repeat the described process with the HWP removed in order to divide out any spectral affects of the FTS setup and normalize the transmission through the HWP. We alternate these sample-out runs with the sample-in runs to suppress effects due to drift in the system. Additionally, we repeat the measurement at various HWP azimuth positions to average over any polarization of the ceramic source induced by the chopper blade or the input mirror.

The result of the FTS measurement is shown by the ``Data'' curve in Figure \ref{fig:hwpBandpass}. We integrate the measured transmission across each PB2 frequency band to obtain the 95/150 GHz HWP transmission values in Table \ref{table:measParams}. To estimate the emissivity, we simulate \cite{tomCode} HWP transmission versus frequency using the dielectric layer parameters in Table \ref{table:measOptics} and isolate the loss due to absorption. The result of the transmission simulation is shown by the ``Theory'' curve in Figure \ref{fig:hwpBandpass}; the simulated transmission is consistent with the measured transmission to within 1$\sigma$ uncertainty. The HWP emissivity at 95/150 GHz is calculated by integrating the simulated absorption of the HWP across each PB2 band. The results of the PB2 HWP emissivity estimate are presented in Table \ref{table:measParams}. 


\begin{table}
	\centering
	\begin{tabu}{| c | c | c | c | c | c |}
	\hline
	Band & Transmission & Emissivity & Mod Efficiency & Phase Diff \\
	\hline
	\hline
	95 GHz & $0.959 \pm 0.014$ & $0.020 \pm 0.009$ & $0.989 \pm 0.005$ & \multirow{2}{*}{$1.3 \pm 0.1$ deg} \\
	\cline{1-4} \cline{6-6}
	150 GHz & $0.941 \pm 0.015$ & $0.032 \pm 0.014$ & $0.984 \pm 0.004$ &  \\
	\hline
	\end{tabu}
\caption{Band-integrated values of key validation parameters across each PB2 frequency channel. We present the measured transmission and estimated emissivity using the result shown in Figure \ref{fig:hwpBandpass} and the measured modulation efficiency and differential phase using the result shown in Figure \ref{fig:polModPlot} \label{table:measParams}}
\end{table}


\subsection{Polarization modulation validation} 
 \label{sec:polValid}

\begin{figure}
\centering
\begin{subfigure}{.44\textwidth}
  	\centering
  	\includegraphics[trim={22.0cm, 1.0cm, 15cm, 1.5cm}, clip, width=\linewidth]{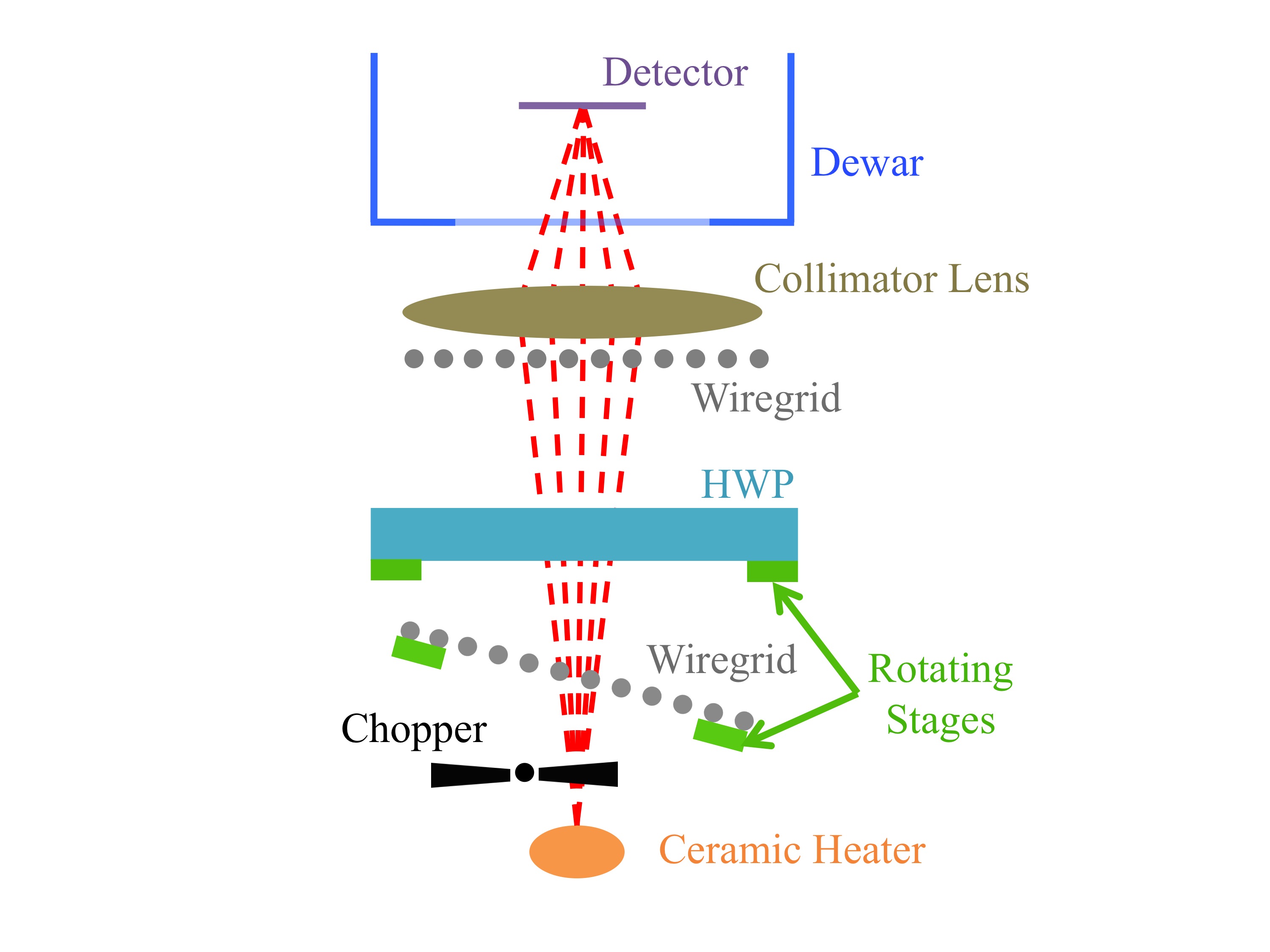}
   	\caption{\label{fig:polModCartoon}}
\end{subfigure}%
\begin{subfigure}{.56\textwidth}
  	\centering
  	\includegraphics[trim={3.0cm, 1.0cm, 3.5cm, 1.5cm}, clip, width=\linewidth]{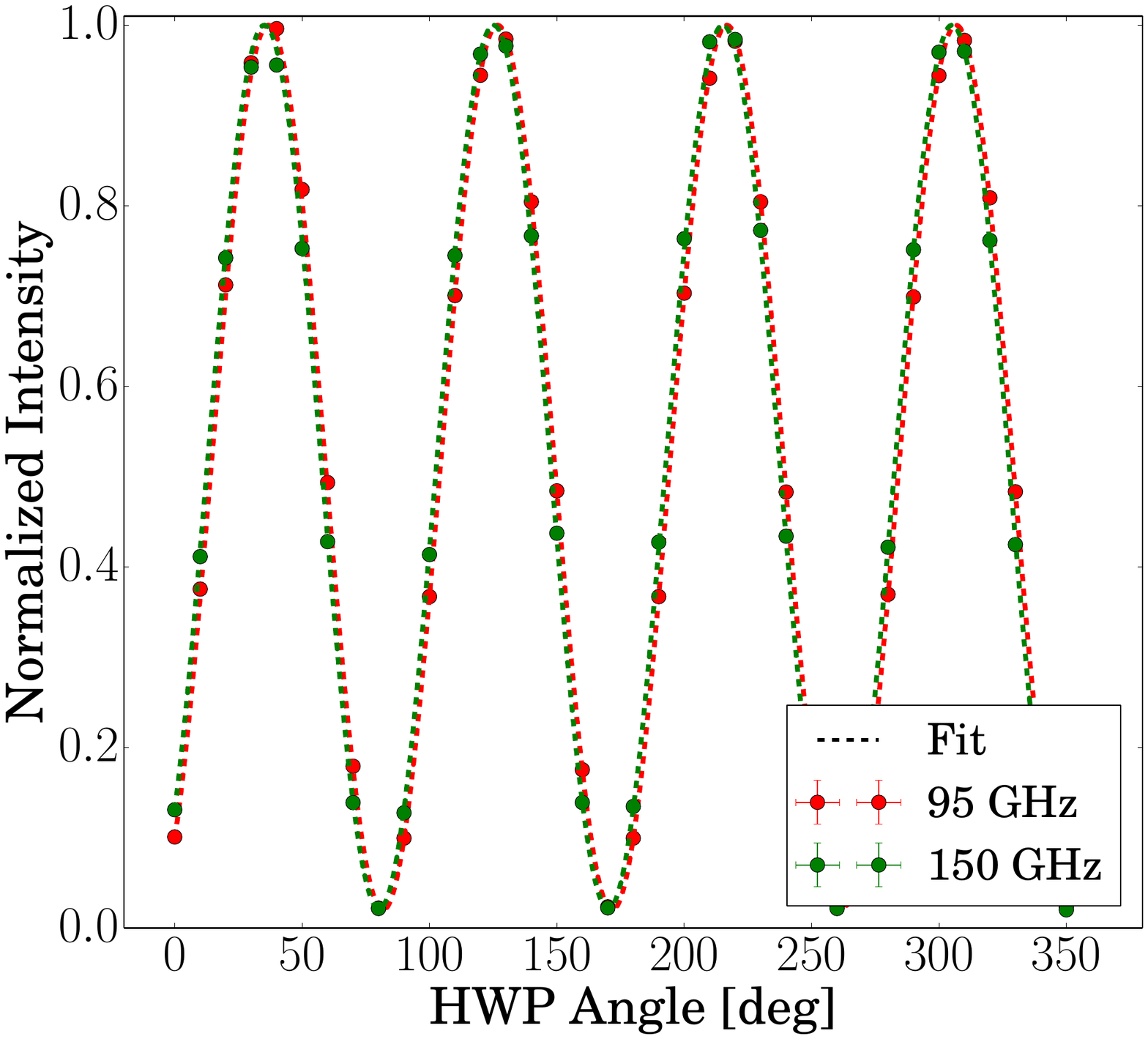}
  	\caption{\label{fig:polModPlot}}
\end{subfigure}
\caption{Measurement of the PB2 HWP polarization performance in the PB2 frequency bands. \ref{fig:polModCartoon} shows a cartoon of the polarized measurement apparatus. The HWP is mounted on a rotating stage to modulate a polarized thermal source with respect to a polarized detector. \ref{fig:polModPlot} shows normalized intensity as a function of HWP angle in each PB2 frequency band. The points are the data and the dotted lines are the fits to Equation \ref{eq:fitPol}. The 95 GHz modulation (in red) is slightly ahead of the 150 GHz modulation (in green). The error bars for the measured data points are small and hence are hidden in this plot. \label{fig:polMod}}
\end{figure}

To validate the HWP polarization performance, we utilize the setup shown schematically in Figure \ref{fig:polModCartoon}. Signal from a temperature-modulated source is polarized with a wiregrid. The wiregrid is tilted to avoid standing waves between it and the chopper. The polarized beam travels through the HWP, is polarized via another wiregrid, and is focused via an UHMWPE collimator lens onto a broadband detector with on-chip filters to define 95 GHz and 150 GHz bands. Even though the detector is polarized, we introduce a wiregrid on the detector-side of the HWP to mitigate the frequency-dependent polarization angle associated with the sinuous antenna \cite{edwards}. The HWP azimuthal angular position is stepped in 10-degree increments and the SQUID output is sent to a lock-in amplifier whose value is integrated for 1.0 s at each HWP orientation.

The results of the measurement are shown in Figure \ref{fig:polModPlot} with a fit to
\begin{equation}
	S_{\mathrm{det}} = \varepsilon \, \cos^{2}\Big[2 (\rho - \phi)\Big] + (1 - \varepsilon) \, ,
\label{eq:fitPol}
\end{equation} 
where $S_{\mathrm{det}}$ is the normalized signal seen by the detector, $\rho$ is the HWP angle, $\phi$ is the HWP modulation phase (Equation \ref{eq:rot}), and $\varepsilon$ is the HWP modulation efficiency (Equation \ref{eq:modEff}).

To characterize the cross polarization inherent to the setup, we remove the HWP and step the azimuthal angle of the tilted wiregrid in 10-degree increments over a 180-degree range. We fit this data to a model given by
\begin{equation}
	S_{\mathrm{det}} = (1 - L) \, \sin^{2}\Big[\rho_{\mathrm{WG}} - \phi_{\mathrm{WG}}\Big] + L \, ,
\label{eq:wg}
\end{equation} 
where $S_{\mathrm{det}}$ is the normalized signal seen by the detector, $\rho_{\mathrm{WG}}$ is the wiregrid azimuthal angle, $\phi_{\mathrm{WG}}$ is some phase, and $L$ is the wiregrid polarization leakage. We find the leakage of the setup to be $<$ 1\% for both the 95 and 150 GHz bands.

After accounting for the leakage due to the setup alone, the estimated polarization efficiency and differential phase for the PB2 bands are shown in Table \ref{table:measParams}. Our measurement is consistent with the theoretically-predicted \cite{tomoTheory} values (see Table \ref{table:plateOrientations}) to within 1$\sigma$ uncertainty associated with the plate orientations and the detector's 95/150 GHz band positions.

We did not measure the absolute phase in this setup, as we did not know the orientation of the HWP with respect to the wiregrids. The differential phase result shows that the polarization angle is controlled to a $\approx$ 1-degree level across the PB2 frequency regime, which is sufficient to validate the HWP prior to polarization angle calibration during integration with the PB2 receiver \cite{keating}.

\subsection{Mechanical validation} 
\label{sec:mechValid}

\begin{figure}
\centering
	\includegraphics[trim={1.5cm, 1.0cm, 1.5cm, 1.5cm}, clip, width=0.7\linewidth]{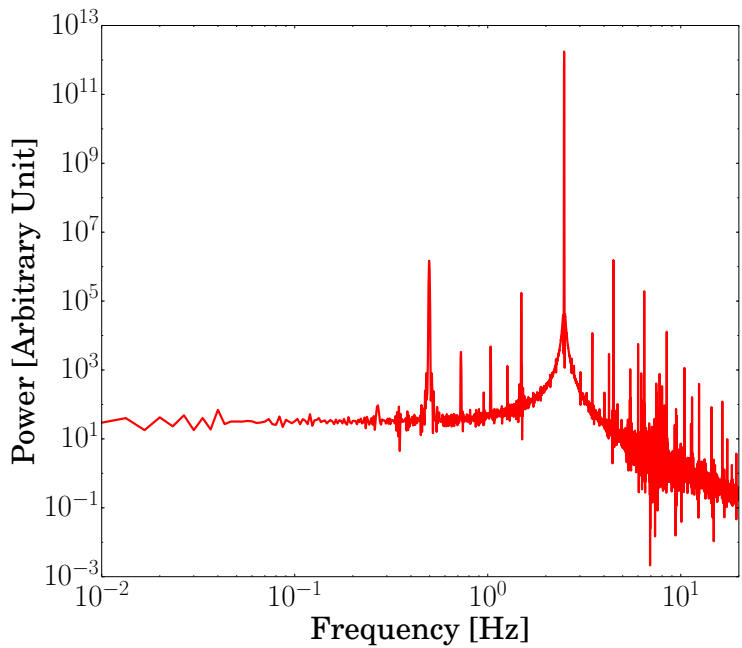}
	\caption{A power spectral density of the HWP rotation at $\approx$ 2.2 Hz over a five-minute period. The excellent rotational stability of the drive train demonstrates the effectiveness of the servo/timing belt system. \label{fig:motorPSD}}
\end{figure}

To validate the HWP rotation mechanism, we must evaluate vibration damping, bearing robustness, rotational stability, and lifetime of the vacuum module. Figure \ref{fig:motorPSD} shows the power spectral density of $\approx$ 2.2 Hz HWP rotation read out at 10 kHz for five minutes. The sharpness of the central peak demonstrates outstanding rotational stability and therefore the effectiveness of the servo/timing belt system. Long-term-rotation, vacuum-module, and vibration-damping tests are all ongoing as we prepare for deployment.

\section{Discussion and Future plans}
\label{sec:future}

We have presented the design and validation of an ambient-temperature achromatic sapphire HWP for observation of the CMB polarization at 95 GHz and 150 GHz. Considering the 1/f noise suppression and systematic error mitigation offered by the implementation of a rapidly-rotating HWP, PB2 will deploy the modulator presented in this proceeding when it begins observations in 2017.

POLARBEAR-2 is the first installment of the Simons Array, an assembly of three telescopes called ``PB2a'', ``PB2b'', and ``PB2c''. PB2a/b will observe at 95/150 GHz, while PB2c will observe at 220/280 GHz \cite{toki}. The ambient-temperature HWP presented in this proceeding will deploy on PB2a.  As we look towards polarization modulation for PB2b/c, we aim to reduce the optical power introduced by the HWP by cooling the modulator to cryogenic temperatures. By moving the HWP into the receiver, we can simultaneously utilize sapphire tan $\delta$'s strong temperature dependence \cite{parshin} and the new location's cryogenic surroundings to dramatically suppress HWP thermal emission and scattering to ambient temperature. A cryogenic HWP prototype is currently under evaluation, and we are in the development stage for the full-scale instrument that will deploy on PB2b/c in 2017/2018.


\acknowledgments 
 
We gratefully acknowledge support for the ambient-temperature HWP development from the National Science Foundation through MSIP Grant \#1440338 and through Advanced Technologies and Instrumentation Grant \#1207892. We acknowledge support for fabrication and testing from the Simons Foundation, Templeton Foundation, and the Gordon and Betty Moore Foundation. This work was supported in part by the U.S. Department of Energy, Office of Science, Office of High Energy Physics, under contract No. DE-AC02-05CH11231 and the Laboratory Directed Research and Development Program at Lawrence Berkeley National Laboratory. MH was supported by
MEXT KAKENHI Grant Number JP15H05891, JSPS KAKENHI Grant Number JP26220709 and the Japan/U.S. Cooperation Program in the field of High Energy Physics.

We also want to thank Nicholas Harrington for his work on the microcontroller control code and encoder readout, Nathan Whitehorn for his work on the infrastructure of the HWP data acquisition, and William Holzapfel for his useful discussions and ideas regarding the vacuum module.


\bibliographystyle{spiebib} 
\bibliography{WHWP_SPIE_BIB} 


\end{document}